\documentclass[runningheads]{llncs}
\usepackage{graphicx}

\usepackage{amssymb}
\usepackage{amsmath}
\usepackage{bbm}
\newtheorem{asm}{Assumption}

\usepackage{algpseudocode}          
\usepackage{algorithm}

\usepackage[font=small]{subcaption}
\begin{document}
\title{Learning Best Response Strategies\\ for Agents in Ad Exchanges \thanks{This work was partially supported from the Greek State Scholarship Foundation by the project ``IKY Scholarship'' from resources of ESF and ESPA.}} 
%
%
\author{Stavros Gerakaris\inst{1} \and
Subramanian Ramamoorthy\inst{2}}
%
%
\institute{School of Informatics, University of Edinburgh \email{stevegerak@gmail.com}\and
School of Informatics, University of Edinburgh
\email{s.ramamoorthy@ed.ac.uk}}
\maketitle              
\begin{abstract}
Ad exchanges are widely used in platforms for online display advertising. Autonomous agents operating in these exchanges must learn policies for interacting profitably with a diverse, continually changing, but unknown market. We consider this problem from the perspective of a publisher, strategically interacting with an advertiser through a posted price mechanism. The learning problem for this agent is made difficult by the fact that information is {\textit{censored}}, i.e., the publisher knows if an impression is sold but no other quantitative information. We address this problem using the Harsanyi-Bellman Ad Hoc Coordination (HBA) algorithm~\cite{albrecht2013game,acr2016aij}, which conceptualises this interaction in terms of a Stochastic Bayesian Game and arrives at optimal actions by best responding with respect to probabilistic beliefs maintained over a candidate set of opponent behaviour profiles. We adapt and apply HBA to the censored information setting of ad exchanges. Also, addressing the case of stochastic opponents, we devise a strategy based on a Kaplan-Meier estimator for opponent modelling. We evaluate the proposed method using simulations wherein we show that HBA-KM achieves substantially better competitive ratio and lower variance of return than baselines, including a Q-learning agent and a UCB-based online learning agent, and comparable to the offline optimal algorithm.

\end{abstract}
\section{Introduction}
Real-time ad exchanges (AdX) have become a common marketplace for online display advertising. These automated transactions take place numerous times a day, when a user visits a web page whose advertising inventory is managed by an AdX. The webpage, which is essentially the \emph{publisher}, communicates a \emph{reserve price} to the ad exchange for the \emph{impression}, which consists of a description of the webpage, of the user and some other relevant content. The ad exchange offers the impression to the bidding agents, or \emph{advertisers}, who compete for it in a second price auction with reserve price, managed by the AdX.

These automated transactions play an important role in the economics of the web, which has meant that advertisers routinely use automated methods to target these impressions to user profiles and characteristics that they are shown. The corresponding situation for publishers appears to be different. As argued in a report from Google~\cite{rtbinsights}, who run one such large exchange, publishers are lagging behind in being able to automate the setting of auction parameters such as reserve price. Furthermore, a nontrivial fraction of ad exchange auctions involve a \emph{single} advertiser~\cite{amin2013learning}. When only one advertiser is involved, the ad exchange auction mechanism becomes a posted price auction between the publisher and the advertiser. 

In this work, we model this interaction and propose the application of novel learning algorithms to address the problem of adapting behaviour within the interaction. We examine the continuous interaction, over a number of rounds, between the advertiser and the publisher through the posted price auction mechanism. There are two key attributes associated with this posted price mechanism; (a) since there are only two agents involved, the observations the publisher makes from the advertiser's bids are \emph{doubly censored} (i.e., the publisher only learns if a bid is successful and does not gain further quantitative knowledge of the advertiser's utilities) and (b) the publisher is facing an \emph{adaptive player} with a number of possible strategies at his disposal. 

Conceptually, the problem faced by the publisher is that of interacting in an {\textit{ad hoc}} manner, with limited prior knowledge of the opponent. Learning in such situations is made difficult by the fact that the open ended nature of the hypothesis space results in unacceptable complexity of learning. We propose that this problem may be addressed by drawing on recent developments in machine learning, which allow tractable learning despite the incompleteness of models. In particular, we use the Harsanyi-Bellman Ad Hoc Coordination (HBA) algorithm~\cite{albrecht2013game,acr2016aij}, which conceptualises the interaction in terms of a space of `types' (or opponent policies), over which the procedure maintains beliefs and uses the beliefs to guide optimal action selection. The attraction of this algorithm is that it can be shown to be optimal even when the hypothesised type space is not exactly correct but only approximates the possible behaviours of the opponent. In this paper, we adapt HBA to the situation where observations are censored, and demonstrate its usefulness in the AdX domain. In addition, addressing the case when opponents are playing essentially randomly (a situation where HBA's belief update process would be inadequate), we propose the use of a Kaplan-Meier estimator to approximate the opponent's stochastic behaviour to choose actions based on that estimate.

We model the interactions between the two agents as a Stochastic Bayesian Game $\Gamma$ of censored observations. The publisher's goal is to maximise his expected revenue over the $T$ rounds of the game. In order to do so he needs to infer the bidding strategy of the advertiser. We define a space of behaviours for the advertiser, including various distributions and adaptive procedures, such as Q-learning and learn-then-bid strategies~\cite{ghosh2009adaptive}. So, a publisher using HBA maintains a belief about the advertiser's behaviour, defined as a probability distribution over this space of types, $\Theta_A$, and plays best response to it. It is worth noting that the \emph{offline optimal} algorithm for the publisher, that serves as an upper bound on the expected revenue of our method, is the strategy that has prior knowledge of the buyer's strategy (something that is unrealistic in practice, but illustrative for algorithm analysis) and plays optimally against it from the first turn of the game. 

There is a substantial body of work in the AdX literature, which focuses on finding a publisher's reserve price policy to optimise his revenue in second price auctions with reserve price~\cite{cesa2013regret,MohriM14}. However, these works, which mainly focus on the theory, often restrict attention to situations such as where an advertiser is only an unknown random distribution (hence, not adaptive), and where the publisher has access to uncensored samples. From a practitioner's perspective, it is interesting to ask if we can go beyond some of these assumptions and devise learning algorithms for the publisher that only uses the censored observations available online (hence making it robust with respect to model mismatch), and allow for more generality in the behaviour of the advertiser, specifically allowing that agent to adapt (which is very realistic in the scenarios we mentioned earlier). In our experiments, we show that the proposed procedure is able to adapt better than baselines such as Q-learning or a UCB-based online learning procedure, and that it approaches the offline optimal benchmark in many situations. In order to understand the behaviour of this algorithm under model mismatch, we also present experiments with an adaptive adversary which is a neural network, looking both at the transient behaviour of HBA when the adversary is actively learning and is non-stationary, and also the case where the adversary is a mixture that is different from any individual element in the type space over which HBA maintains beliefs.

\section{Related Work}

Much has been written about ad display and sponsored search auctions, for each participating agent of the auction, either as publisher or advertiser. A key paper in the area of ad exchanges is that of Muthukrishnan \cite{muthukrishnan2009ad}, who laid out several research directions in this domain, informed by exposure to the practice in this domain.

More specific related work, viewing the problem from the perspective of the publisher, are the following. Mohri and Medina~\cite{MohriM14} discuss selecting the reserve price to optimise revenue in a second price auction with reserve price. They consider a supervised learning setting and assume access to a training sample of uncensored historical data. A similar formulation of revenue is seen in the work of Cesa-Bianchi et al.~\cite{cesa2013regret}, where the authors assume no historical data, but they get direct observations based on the assumption that every bidder in this market draws his valuation from the same fixed and unknown distribution $D$. Then they proceed by showing a regret minimisation algorithm, achieving sublinear regret. In other recent work, Amin et al. \cite{amin2013learning} define the notion of the \emph{strategic regret} and present no-regret algorithms with respect to that notion. Finally, Huang et al.~\cite{HuangMR14} study the problem of setting a price for a potential buyer with a valuation drawn from an unknown distribution $D$ and prove tight sample complexity bounds for regular distributions, bounded-support distributions, and a wide class of irregular distributions. This work is preceded by Cole and Roughgarden~\cite{cole2014sample}, who also analyse sample complexity of revenue maximisation, this time as a function of the number of bidders in an auction.

From the advertiser's perspective, Amin et al.~\cite{amin2012budget} study budget optimisation for sponsored search auctions. The authors cast the problem of budget optimisation as a Markov Decision Process (MDP) with censored observations and propose a learning algorithm based on Kaplan-Meier estimators. The authors also perform a large scale empirical demonstration on auction data from Microsoft, in order to demonstrate that their algorithm is extremely effective in practice. Another take on the advertiser's optimisation problem is the one by Ghosh et al.~\cite{ghosh2009adaptive} who study the design of a bidding agent in a marketplace for displaying ads. They provide algorithms and performance guarantees for both settings, while also experimentally evaluating their performance on a fitted log-normal distribution from data observed from the Right Media Exchange.

Another literature that is closely relevant pertains to learning to interact in multiagent domains, with limited or no prior coordination. Related work includes~\cite{albrecht2013game,barrett2011empirical,stone2010ad}. A key concept arising from this literature is that of modelling the opponent in terms of a hypothesis space of policies, that in a certain sense approximate the space from which that agent herself chooses the true policy. 

\section{Model for the Publisher in an Ad Exchange}

We start by presenting our model of how we conceptualise this interaction with the (model of an) advertiser, agent $A$, as a Stochastic Bayesian Game. The advertiser is characterised by a discrete state space $\mathcal{S}$, defined by his own budget $B_A$ and the auction round $t$, $s^t = \{B_A^t, t\}$. He has a set of actions $\mathcal{A}_A =\{ 0, \dots, v_{max}\}$ which are the possible prices he can bid for an impression (his valuation vector) and his strategy is chosen from a well-defined type space $\Theta_A$. A payoff function $u_{A} : \mathcal{S} \times \mathcal{A} \times \Theta \mapsto \mathbb{R}$ maps his state, type and actions to specific payoff and a strategy $\pi_A : \mathbb{H} \times \mathcal{A}_A \times \Theta_A \mapsto [0,1]$ defines a probability vector over his possible actions. The history vector $\mathbb{H}$ contains all histories $\mathbb{H}^t = \langle s_0, a_0, s_1, a_1, \dots,s_{t-1}, a_{t-1}, s^t \rangle$.

We realise the interaction between the advertiser and the publisher as a Stochastic Bayesian Game $\Gamma$ of censored information between two players. The game $\Gamma$ consists of:

 \begin{itemize}
  \item An advertiser $A$ and a publisher $P$
  \item State space $\mathcal{S}$, action space $\mathcal{A} = \{\mathcal{A}_A, \mathcal{A}_P \}$, type space $\Theta = \{ \Theta_A, \Theta_P \}$ 
  \item Transition function $ T : \mathcal{S} \times \mathcal{A} \times \mathcal{S} \mapsto [0,1]$ 
  \item $\Gamma$  starts at time $t = 0$ and state $s^0$. At each time step $t \in 0, 1, 2, \dots$:
  \begin{enumerate}
  	\item An impression $i^t$ arrives
  	\item The advertiser chooses his action (bid) $ b^t = a_A \in \mathcal{A}_A$ with probability $\pi_A(H^t, a_A^t, \theta_A^t)$ 
	\item The publisher chooses his action (reserve price) $r^t= a_P \in \mathcal{A}_P$ with probability $\pi_P(H^t, a_P^t, \theta_P^t)$ 
	\item The game transitions into state $s^{t+1} \in S$ with probability $T(s^t, a^t, s^{t+1})$
	\item If $b^t \geq r^t$ the impression is sold at price $r^t$; otherwise the impression doesn't get sold
  \item The immediate payoff of the advertiser $A$ is $u_A^t = b^t - r^t$ 
  \item The immediate payoff of the publisher $P$ is $u_P^t = r^t$
  \end{enumerate}
  \item The process is repeated until a terminal state $s^t$ is reached
  \end{itemize}

In this problem setting, the publisher does not have knowledge of the advertiser's individual strategy $\theta_A \in \Theta_A$, only of his type space $\Theta_A$ which is the set of all of his possible strategies. He needs to infer that strategy $\theta_A$ from the censored observations he makes at each auction round in order to play his best response strategy against it.  As mentioned earlier, we utilise the Harsanyi-Bellman Ad Hoc Coordination Algorithm~\cite{albrecht2013game,acr2016aij}. HBA, and adapt it to the setting of AdX. The main steps of this procedure are as described in algorithm \ref{hba}. A key adaptation from the formulation in~\cite{albrecht2013game,acr2016aij} is the incorporation of an estimate of the opponent's actions and to allow for the KM estimator (to be explained in more detail in the next section) for the case of a randomised adversary.

We make the following assumptions for our setting: 

\begin{asm}
We control the publisher $P$ and choose his strategy $\pi_P$. $P$ has a single type $\theta_P$ known to us.
\end{asm}

\begin{asm}
Given a stochastic game $\Gamma$ we assume all the elements of $\Gamma$, except of the type $\theta_A$ of the opponent, is common knowledge.
\end{asm}

\begin{asm}
We only have partial observability of states and actions.
\end{asm}

\begin{algorithm}
\caption{HBA Censored}\label{hba}
\begin{algorithmic}[1]
\Require SBG $\Gamma$, player $P$, user defined type spaces $\Theta^*_A$, history $H^t$, discount factor $0 \leq \gamma \leq 1$
\Ensure Action probabilities $\pi_P(H^t, r)$
\ForAll{$\theta^*_A \in \Theta^*_A$}
	\State Compute $Pr(\theta^*_A | H^t) = \frac{L(H^t | \theta^*_A)P(\theta^*_A)}{\sum_{\hat{\theta}^*_A \in \Theta^*_A}L(H^t | \hat{\theta}^*_A)P(\hat{\theta}^*_A)}$ \label{ln:post}
\EndFor
\ForAll{$r \in \mathbf{v}$}
	\State Compute expected payoff $E_{s^t}^{r}(H^t)$ as follows:
	$$
	E_{s}^{r}(\hat{H}) \leftarrow \sum_{\theta^*_A \in \Theta^*_A}Pr(\theta^*_A | H^t) \sum_{b \in \mathbf{v}} Q_s^{r}(\hat{H})  \pi_A (\hat{H}, b, \theta^*_A )
	$$
	$$
	Q_s^{r}(\hat{H})  \leftarrow \sum_{s' \in S} T(s, r, s') \Big[ u(s, r) + \gamma \max_{r'} E_{s'}^{r'}(\langle\hat{H}, r, s'\rangle) \Big]
	$$ \label{ln:em}
\EndFor
\If{$\arg\max_{\theta_A \in \Theta_A} \in \Theta_A^{\text{Random}}$}\label{ln:choice}
	\State $\pi_P(H^t, r^*) = 1$, where $r^* = \textsc{Random KM} (k,l, k_c)$
\Else
	\State Distribute $\pi_P(H^t, \cdot)$ uniformly over $\arg\max_{r} E_{s^t}^r(H^t)$
\EndIf
\end{algorithmic}
\end{algorithm}

In the HBA algorithm, which maintains a posterior probability of an agent being a specific type based on observing the history of actions, the action is selected by determining a best response, within the game $\Gamma$, which implicitly uses in the value calculations Q-values based on the Bellman optimality principle.

The posterior probability of an agent being a specific type is calculated with the use of sum posterior, defined in \cite{albrecht2014convergence} as:
$$
L(H^t) = \sum_{\tau = 0}^{t-1} \pi_j(H^{\tau}, a_j^t,\theta_j)
$$


By the term \emph{censored observations} we refer to the type of the information perceived by the publisher. As is the case in posted price auctions, the publisher only gets to observe the outcome of any round $t$ of a sequential auction, which is if he sold the impression or not, but he doesn't get to observe the bid that actually won; he only knows that this bid is greater or equal than the reserve price he specified ($b^t \geq r^t$). Otherwise, he knows that this bid was strictly less than his reserve price ($b^t < r^t$). 

\subsection{HBA Types (Advertiser's Strategies)}\label{sec:hbatypes}

In this section, we define the hypothesised type space {$\Theta_A$} of the advertiser. The first two strategies don't involve an adaptive component, so they are in a sense naive. However, there are works \cite{pin2011stochastic} that suggest that this kind of bidding can often be found in real world auctions. The rest of the strategies of the advertiser that we specify, are well studied learning models that involve distinct learning and strategy components. This set is chosen to capture the diversity of potential types of behaviour of the unknown adversary.

We also present the best response strategies of the publisher to each of the respective strategies of the advertiser, under the assumption that all the private information of the advertiser is known by the publisher. These best response strategies consist a set of \emph{offline optimal} benchmarks, that will serve as an upper bound on the revenue of our method, which assumes no private knowledge other than the type space of the advertiser. 

\paragraph{Greedy Strategy}
Advertiser's greedy policy, given his action space $\mathcal{A_A} = \{ 0, \dots, v_{\max}\}$, is to always to bid his maximum value for the impression. 
$$
\pi_{A}^{\text{greedy}}(a_A = v_{\max}) = 1
$$
One can see that publisher's best response policy is to simply match his maximum value and offer it as a reserve price in every turn of the game. 
$$
\pi_P^{*}(a_P =v_{\max} | \pi_{A}^{\text{greedy}}(a_A \in \mathcal{A}_A)) = 1
$$

\paragraph{Random Strategy}
In the second strategy, the advertiser places a bid i.i.d. from a fixed distribution over his value vector. We use several random distributions, such as the uniform, the normal, the logistic, the log-normal and the exponential. 

The best response strategy against a random advertiser, with the distribution $\mathcal{D}$ over $\mathbf{v}$ known by the publisher, is the reserve price that maximises the publisher's expected revenue.  
$$
\pi_{P}^{*}(a_P= r^* | \pi_{A}^{\text{random}}(b \sim_{\mathcal{D}} \mathbf{v})) = 1,  r^* = \arg\max_{r \in \mathbf{v}} r T_{\mathcal{D}}(r)
$$
where $T_{\mathcal{D}}(r)$ denotes the tail probability of the distribution $\mathcal{D}$ for the value $r$.

\paragraph{Learn Then Bid Strategy}
The next adaptive bidding strategy of the advertiser is given in the work of Ghosh et al. \cite{ghosh2009adaptive}, where the advertiser chooses to opt out for a specified number of $m$ out of $n$ rounds in order to observe the prices of the reserve and then, based on his observations, decides between the price $P_m^*$ that guarantees, in expectation, the target fraction $f$ of impressions he sets, and the price $Z_m^*$ satisfying the maximum amount he wants to spend. 

The best response strategy against an advertiser playing Learn Then Bid strategy, with the parameters of the Learn Then Bid algorithm known by the publisher, is the following. 

\begin{enumerate}
\item Find the maximum value of the price that satisfies the target of his campaign reach, times the probability of him playing that price. 
$$P_m^* = \arg\max P_m \frac{fn}{(n-m) \mathcal{P}_m(P_m)}$$
\item Find the maximum value of the price that satisfies the advertiser's target spent.
$$Z_m^* = \sup{\{z : \mathcal{P}_m(z) \geq \frac{fn}{n-m}\}}$$
\end{enumerate}

$\mathcal{P}_m(x)$ is the estimated distribution of the market from the advertiser after the learning phase. The optimal policy for the publisher is to exhaust the advertiser's budget, by deterministically selecting the maximum of those two prices.
$$
\pi_P^{*}(a_P =\max{\{P_m^*, Z_m^*\}} | \pi_{A}^{\text{Learn Then Bid}}(a_A \in \mathcal{A}_A)) = 1
$$

\paragraph{Multi Armed Bandits Strategy}
Another strategy we use for the advertiser is the well known UCB algorithm \cite{auer2002finite}. We implement it using a $\epsilon$-greedy action selection policy. Publisher's optimal policy is to offer the maximum value, as a reserve price in every turn of the game. 
$$
\pi_P^{*}(a_P =v_{\max}, \cdot | \pi_{A}^{\text{UCB}}(a_A \in \mathcal{A}_A)) = 1
$$
It is not hard to see that any traditional \emph{no-regret} strategy is easily manipulable, therefore inadequate for this interactive problem setting, something also highlighted in \cite{amin2013learning}.

\paragraph{Q-learning Strategy}
The last algorithm in the type space of the advertiser is the well known Q-learning algorithm \cite{watkins1992q}. We implement it using a soft-max action selection policy. The states for the advertiser are the different levels of his remaining budget and the action space is defined by his value vector.

Publisher's optimal policy, similar to when he faces a random distribution, is finding the reserve price that maximises his expected revenue w.r.t. the Boltzmann distribution produced by the Q-values, which dictates the soft-max action selection.
$$
\pi_P^{*}(r^* | \pi_{A}^{\text{Q-learning}}(s, r \in \mathcal{A}_A)) = 1
$$
where,
$$
r^* = \arg\max_{r \in \mathcal{A}_P} r \sum_{r' = r}^{v_{max}} \frac{e^{Q(s, r') / \tau}}{\sum_{a'_A \in \mathcal{A}_A}e^{Q(s, a'_A) / \tau}}
$$

\subsection{HBA Beliefs and Best Responses}

Over the recently introduced types, HBA maintains and updates beliefs, that will determine action selection at each step. In step \ref{ln:post} of Algorithm \ref{hba}, HBA keeps a posterior belief over types, $Pr(\theta^*_A | H^t)$ by keeping track of the sequence of actions of the opposing player and calculating the probability that these actions come from a specific type. Afterwards, in step \ref{ln:em}, it computes the Q-values of every possible action at this state $Q_s^{r}(\hat{H})$ and, following this, it calculates the expected revenue $E_s^{r}(\hat{H})$ of every action based on the posterior over types it maintains and on the Q-values it has just computed. In the final step \ref{ln:choice}, depending on whether it recognises a stochastic opponent, $\theta_A \in \Theta_A^{\text{Random}}$, or a deterministic one, HBA decides between calling a procedure designed specifically for random opponents (discussed in Section \ref{sec:rkm}) and playing a single price as a best response.

By exchanging between iterations of these two procedures, the posterior belief calculation and the Q-values computation followed by a single expectation maximisation step, HBA succeeds in modelling in a dynamic fashion the opposing agent, while also plays optimally, in expectation, against her at each step. 

\subsection{HBA Censored}

As mentioned earlier, accommodating censored observations requires estimating actions that can be used for belief updating. There are two specific values that are needed for such an estimate. The first one concerns the probability of the last observed action, conditional to a player being of a specific type. This probability is used to calculate the posterior of the opposing agent's type, according to the Bayes rule. In our setting, where the observations are censored, we estimate this value by using structural characteristics of the distribution he plays. For instance, let $\mathcal{Q}$ be the distribution associated with the Q-values for a Q-Learning advertiser. If the publisher sells the impression at price $r$, he doesn't observe the bid, but he can update the probability, conditional on his opponent's type:
$$
Pr (a^{t-1}_{A} | \theta_A^{\text{Q-learning}}) = T_{\mathcal{Q}}(r)
$$

The second one concerns the computation of the HBA's own Q-values in the expectation maximisation step. Recall from Algorithm \ref{hba} that we compute the Q-values by computing every possible outcome in expectation, 
$$
Q_s^{v}(\hat{H})  \leftarrow Q_s^{ v}(\hat{H})  + \alpha [v + \gamma \max_{r'} Q_{s'}^{ v' }(\hat{H})  - Q_s^v(\hat{H})],
$$

where $v$ denotes the utility of the previous step:

\begin{align*}
v =
 \begin{cases}
   v, & \text{if } v \leq r \text{ and the impression was sold}, \\ 
   0, & \text{if } v \geq r \text{ and the impression was not sold},\\
  v  T_{\mathcal{Q}}(v), & \text{otherwise.}
  \end{cases}
\end{align*}
Again here we utilise the tail probability of the distribution in order to estimate the required Q-values from the censored observations. 

With this, we achieve performance close to the offline optimal metric, against advertising strategies that play a single price or that are converging to a price, i.e. Greedy, Learn Then Bid or UCB. 

From the defined type space $\Theta_A$ we see cases where the output of our opponent's algorithm is randomised, either according to a fixed random distribution or a dynamic one, in the case of the Q-learning algorithm. It is known from the theory underpinning HBA \cite{albrecht2014convergence} that this case requires different treatment.

\subsection{KM Estimator for Stochastic Opponents}\label{sec:rkm}

We now present an approach based on the Kaplan-Meier estimator \cite{kaplan1958nonparametric}, for estimating distributions from censored samples. When HBA recognises a randomised opponent, we let this algorithm decide both the query values and the optimal reserve price. KM estimator is a powerful tool for approximating distributions based on censored samples and has found use in both e-commerce \cite{amin2012budget} and financial applications \cite{ganchev2010censored}. 

The {Random KM} algorithm uses a few simple ideas from random sampling and the Kaplan-Meier estimator. We start by scanning the support of the distribution for potential candidate values and for every candidate we query for a sufficient number of times, in order to have a good estimation. 

\begin{algorithm}[!ht]
\caption{Random Querying - KM}
\begin{algorithmic}[1]
\Require Distribution $q$ to make CDF queries
\Ensure Optimal reserve price $r_q^*$
\Procedure{Random KM}{$k, l, k_c$}
\For{$t = 1,\dots,k$}
	\State Set the reserve price $r^t$ uniformly at random.
	\If {$b^t \geq r^t$}
		\State $\forall x$ such that $x \leq r^t: R^t(x) \leftarrow R^t(x) + 1$
	\Else
		\State $\forall x$ such that $x \geq r^t: L^t(x) \leftarrow L^t(x) + 1$
	\EndIf
\EndFor
\State $\forall x \in \text{support}: \hat{T}_q(x) = \frac{R^k(x)}{R^k(x)+L^k(x)}$
\State Compute $r_q^*= \arg\max_{r \in \text{support}}r \hat{T}_q(r)$  
\State Create the list $\text{candidates} = [r_q^*-l, \dots, r_q^*+l]$
\ForAll{$c \in \text{candidates}$}
	\State Set the reserve price $r^t = c$ for $k_c$ steps.
	\State Keep counter $R_{c}^t = \sum_{\tau = 1}^t \mathbbm{1}_{\{b^{\tau} \geq r^{\tau}\}}$
	\State Update: 
	$$\hat{T}_q(c) = \frac{R^{k_c}_c}{R^{k_c}_c+L^{k_c}_c} $$
\EndFor
\State $r_q^*= \arg\max_{r \in \text{candidates}}r \hat{T}_q(r)$ 
\EndProcedure
\end{algorithmic}
\label{alg:rkm}
\end{algorithm}

Essentially the {Random KM} algorithm works in two steps. In the first step, it queries for $k$ steps over all the support of the opponent's possible values and makes a loose estimation of each value's tail probability by calculating the fraction of \emph{right censored} observations ($R^t(x)$, times the impression gets sold and the value $x \leq r$ is less or equal than the reserve price), to the sum of right and \emph{left censored} observations.

In the second step it isolates the candidate values which are the most probable to generate the most revenue, and queries each of them for a number of $k_c$ steps, which Kaplan-Meier dictates, in order to get a precise approximation of their tail probabilities. Using the estimated tail probabilities, it calculates the price that maximises the expected payoff and returns it.

\begin{figure}[h!]
        \centering
        \begin{subfigure}[b]{0.24\textwidth}
                \includegraphics[width=\textwidth]{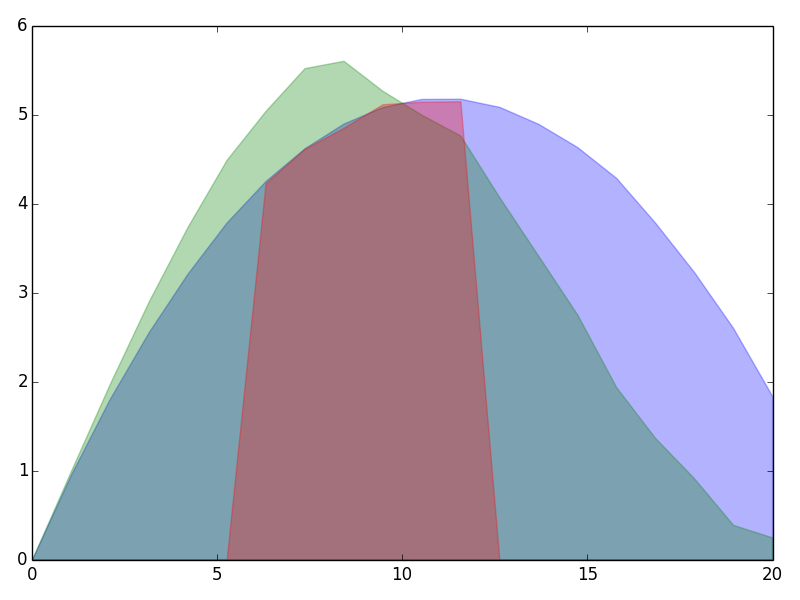}
                \caption{Uniform}
        \end{subfigure}
        \begin{subfigure}[b]{0.24\textwidth}
                \includegraphics[width=\textwidth]{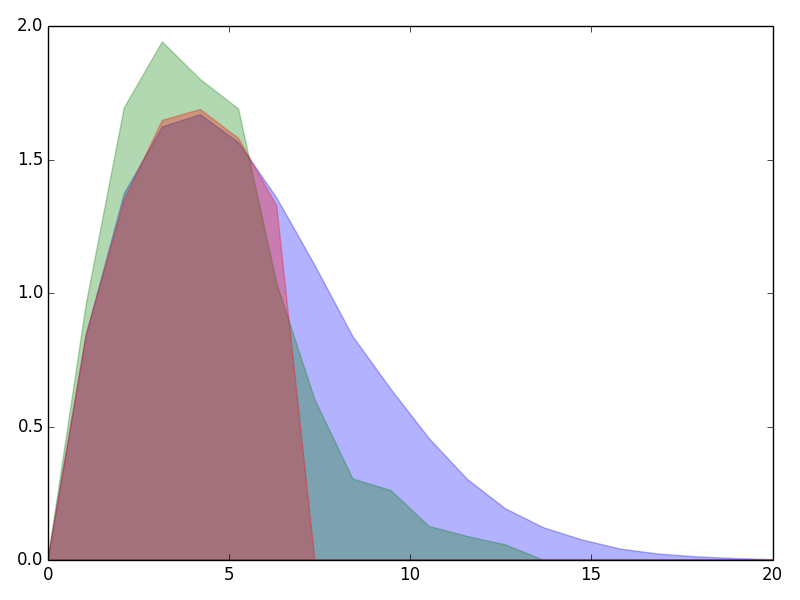}
                \caption{Normal}
        \end{subfigure} 
        \begin{subfigure}[b]{0.24\textwidth}
                \includegraphics[width=\textwidth]{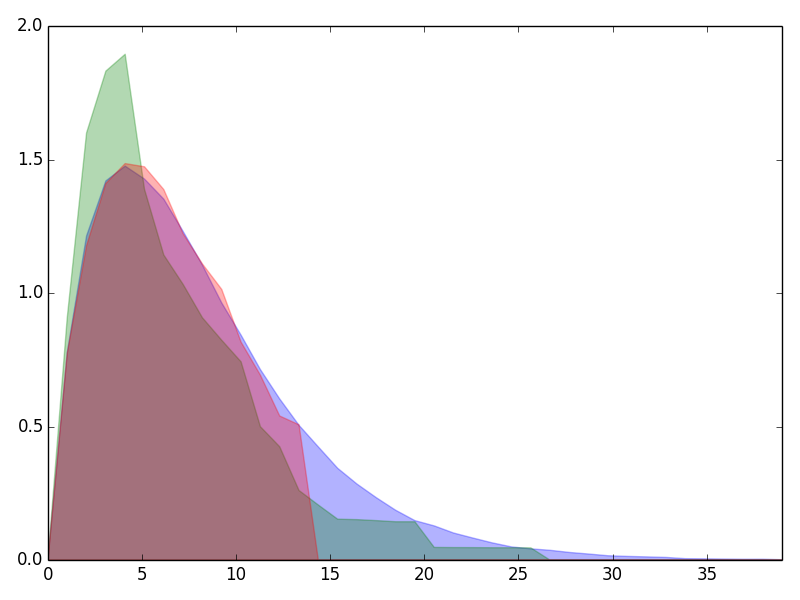}
                \caption{Exponential}
        \end{subfigure} 
        \begin{subfigure}[b]{0.24\textwidth}
                \includegraphics[width=\textwidth]{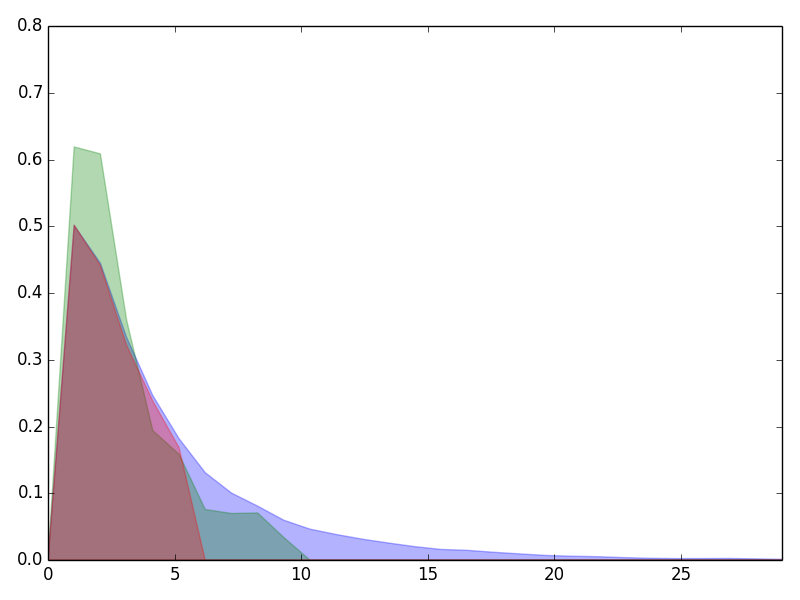}
                \caption{Log-normal}
        \end{subfigure} 
\caption{Random KM estimation of the empirical payoff functions of various random distributions. The loose estimation is the random querying step (in green), followed by the precise KM step (in red).}\label{fig:km}
\end{figure}

In Figure \ref{fig:km} we can see how these two steps are implemented in practice. The green area denotes the estimation of the revenue function, denoted by the blue area, during the first step (random querying). Similarly, the red area denotes the approximation of the expected revenue after implementing the KM estimator on the second step, for a selected number of candidate values. The precise approximation KM provides us, allows for optimal, in expectation, action selection against stochastic opponents. 

\section{Experimental Results}

\subsection{Agents in the type space $\Theta_A$}
The AdX domain is of significant commercial importance. However, this also means that obtaining real world data from live auctions is difficult. We conduct empirical studies using a domain that captures many aspects of this domain. The domain is that of the \emph{Trading Agents Competition (TAC) AdX '15} \cite{schain2013ad}, which simulate an Ad Exchange game. We use our own implementation of these specifications. In particular we have a setting of $60$ days, with $1000$ impression opportunities each day, specified daily Budget $B$ and Campaign Reach $C_R$ for the advertiser and defined advertiser's type space: $\Theta_A =$ [Random, Greedy, LTB, UCB, Q-learn]
  
We use three basic benchmarks for the evaluation of our HBA-KM algorithm. 
\begin{enumerate}
  \item The Offline Optimal algorithm that knows the true type $\theta_A$ of the advertiser \emph{a priori} and decides the optimal policy $\pi^*_P$ as his best response.
  \item The Q-learning algorithm, a well known reinforcement learning technique.  
  \item The UCB algorithm, a MAB technique. 
 \end{enumerate}

The Offline Optimal algorithm realises the best response strategies, discussed in Section \ref{sec:hbatypes}, and serves as an upper bound for our method. We also choose a Q-learning agent and a UCB-based online learning agent as our baselines, since, given the stochastic game formulation of the problem, one may hope to solve it using techniques from reinforcement learning. Q-learning with soft-max, and UCB with $\epsilon$-greedy action selection, are two of the simplest algorithms for reinforcement learning, giving good results in a wide spectrum of applications.

We use different parameters for each of the strategies in the advertiser's type space. For the comparative evaluation of our algorithm, against the specified benchmarks, we consider metrics, such as the \emph{revenue} of our algorithm and the \emph{standard deviation} to quantify our agent's payoff variation between consecutive games.

The parameter settings for our experiments were chosen for the opposing agents, in a way to demonstrate that our results hold, for every way one could distribute the probability mass over the value vector for the impressions ($\mathbf{v} = [0,1]$), according to a specific random distribution or an adaptive strategy. The exact parameters that were used for all the simulations follow.  

\begin{enumerate}
\item For the \emph{randomised} strategies:
$\mathcal{U}\{a, b\}$ for the uniform distribution, with $a = 0$ and $b = \{0.5, \dots, 1\}$.
$\mathcal{N}(\mu, \sigma^2)$ for the normal distribution, with $\mu = \{0.25, \dots, 0.65\}$ and $\sigma^2 = \{2 \times 10^{-6}, \dots, 6 \times 10^{-6} \}$.
 $\ln \mathcal{N}(\mu, \sigma^2)$ for the log-normal distribution, with $\mu = \{-7, \dots, -5.4\}$ and $\sigma = \{ 0.5, \dots, 1\}$.
$f(x; \beta) $ for the exponential distribution, with $\beta = \{1/900, \dots, 1/500\}$.
\item For the \emph{deterministic} and \emph{adaptive} strategies:
For the \emph{Greedy} strategy, we set the bid to be $v_{max} = \{0.5, \dots, 1\}$.
For the \emph{Learn-Then-Bid} strategy, we set the exploration length to be $m = \{ 20, \dots, 200\}$ and the target fraction $f = \{0.3, \dots, 0.7\}$.
For the \emph{UCB} strategy, we set the exploration step $k = \{ 20, \dots, 200\}$ and the exploitation with probability $1-\epsilon$, where $\epsilon = \{ 0.01, \dots, 0.30\}$.
For the \emph{Q-learn} strategy we set the learning rate to be $\alpha = \{0.1, \dots, 0.3\}$, the discount factor $\gamma = \{0.80, \dots, 0.99 \}$ and the temperature of the soft-max selection policy $\tau = \{100, \dots, 1000 \}$
\end{enumerate}

For every opposing strategy, we consider the cartesian product of the parameters we specified and the results that follow are averaged over every possible run using these parameters, across 100 simulations for each individual opponent.

\begin{figure}[!ht]
    \centering
    \begin{subfigure}[b]{0.45\textwidth}
            \includegraphics[width=\textwidth]{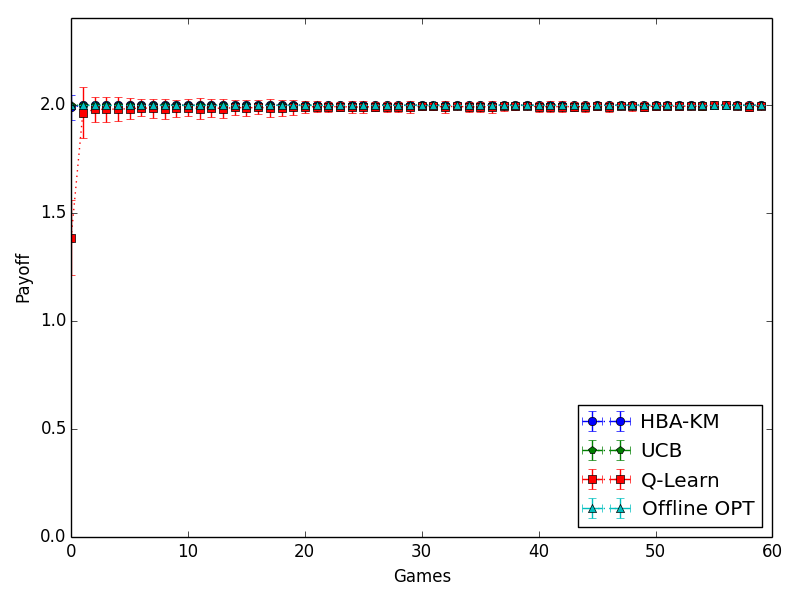}
            \caption{Greedy strategy}
    \end{subfigure}%
    ~ 
    \begin{subfigure}[b]{0.45\textwidth}
            \includegraphics[width=\textwidth]{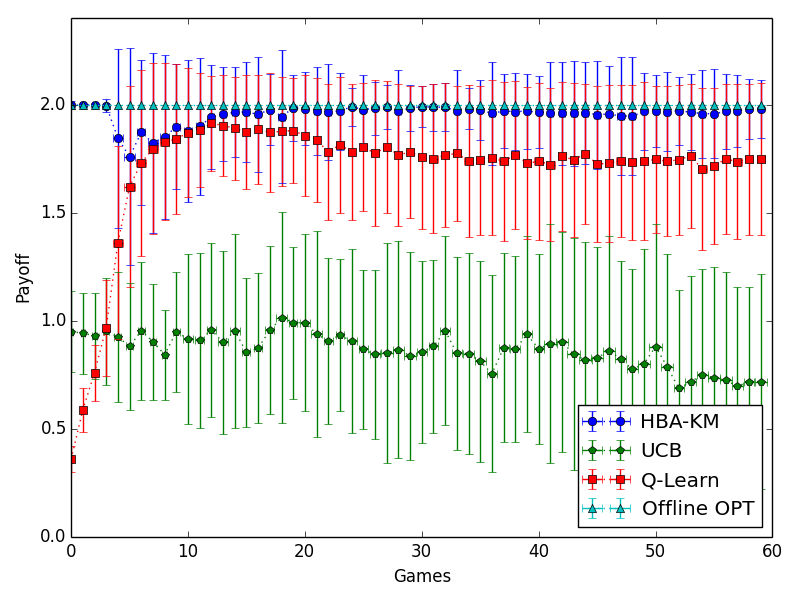}
            \caption{Learn Then Bid strategy}
    \end{subfigure}
    \\
    ~ 
    \begin{subfigure}[b]{0.45\textwidth}
            \includegraphics[width=\textwidth]{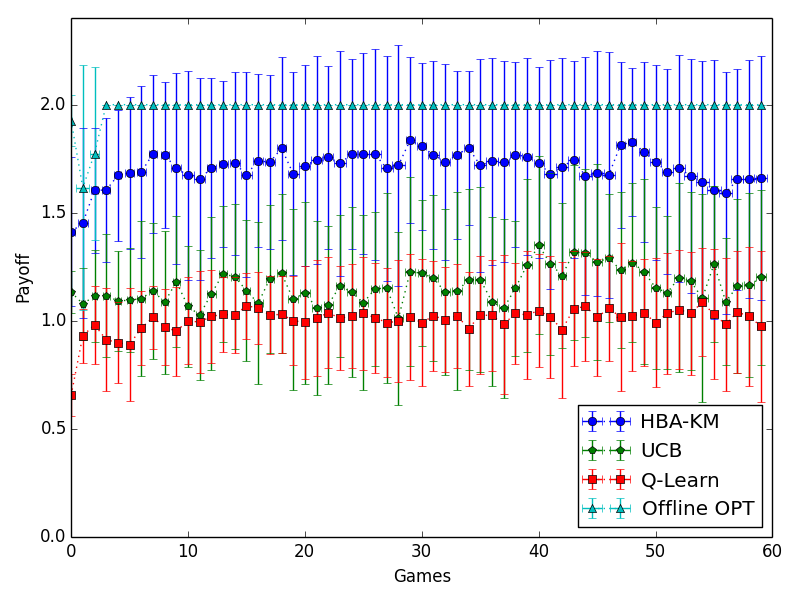}
            \caption{UCB strategy}
    \end{subfigure}
    ~
    \begin{subfigure}[b]{0.45\textwidth}
            \includegraphics[width=\textwidth]{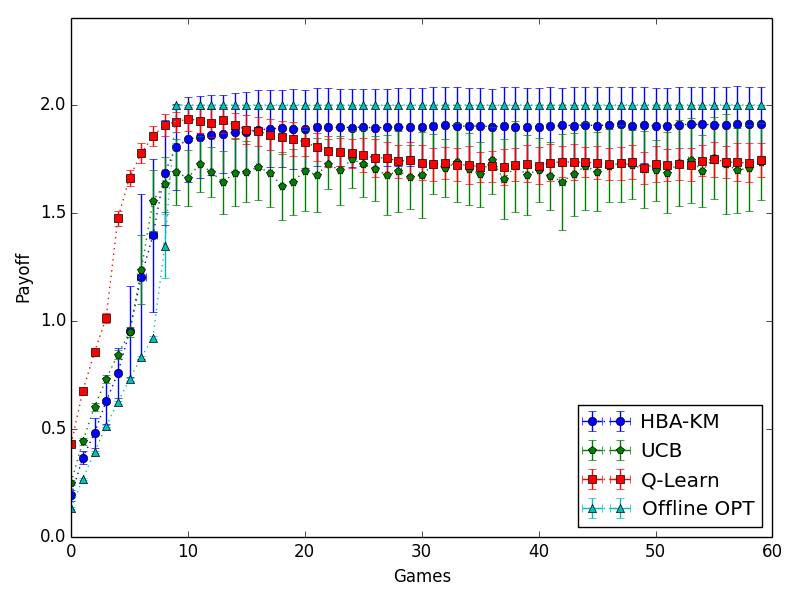}
            \caption{Q-learn strategy}
    \end{subfigure}
\caption{Revenue comparison and one standard deviation against the adaptive strategies.}\label{fig:det}
\end{figure}

In Figure \ref{fig:det}, we can see the HBA-KM's performance against the deterministic and adaptive strategies of the advertiser. The performance is close to the offline optimal benchmark and outperforms the other two baselines.

\begin{figure}[!ht]
\centering
\begin{subfigure}[b]{0.45\textwidth}
        \includegraphics[width=\textwidth]{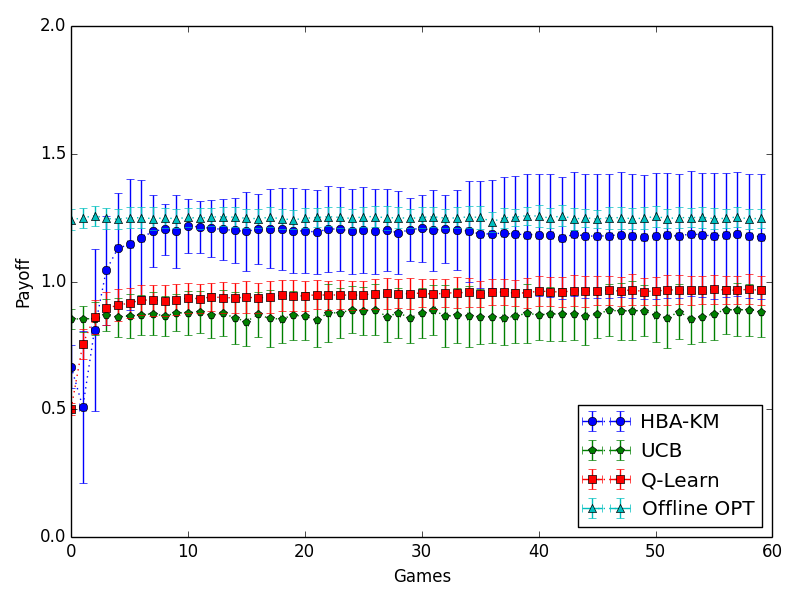}
        \caption{Uniform distribution}
\end{subfigure}
~
\begin{subfigure}[b]{0.45\textwidth}
        \includegraphics[width=\textwidth]{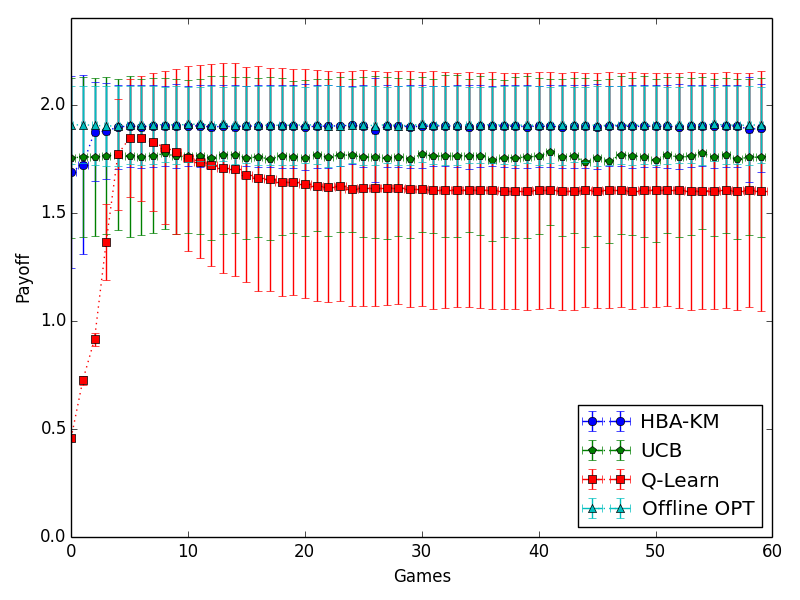}
        \caption{Normal distribution}
\end{subfigure}
~
\begin{subfigure}[b]{0.45\textwidth}
        \includegraphics[width=\textwidth]{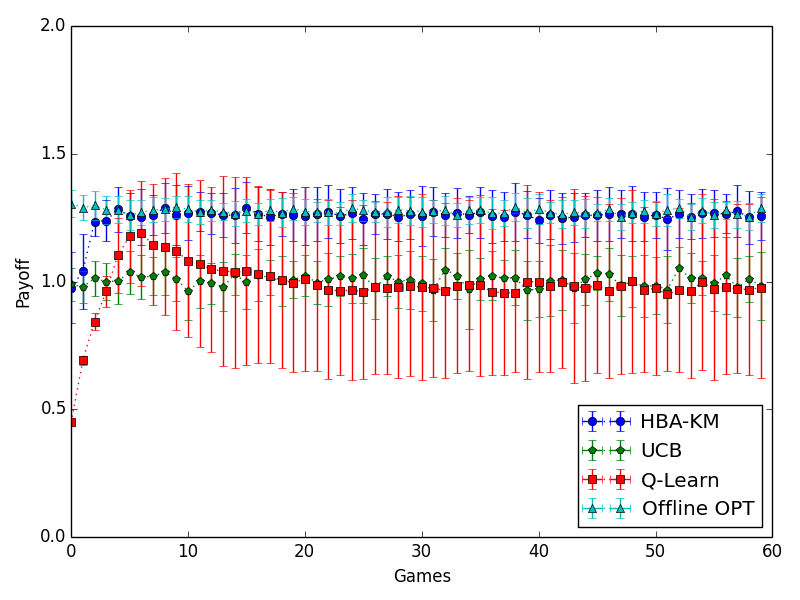}
        \caption{Log-normal distribution}
\end{subfigure}
~
\begin{subfigure}[b]{0.45\textwidth}
        \includegraphics[width=\textwidth]{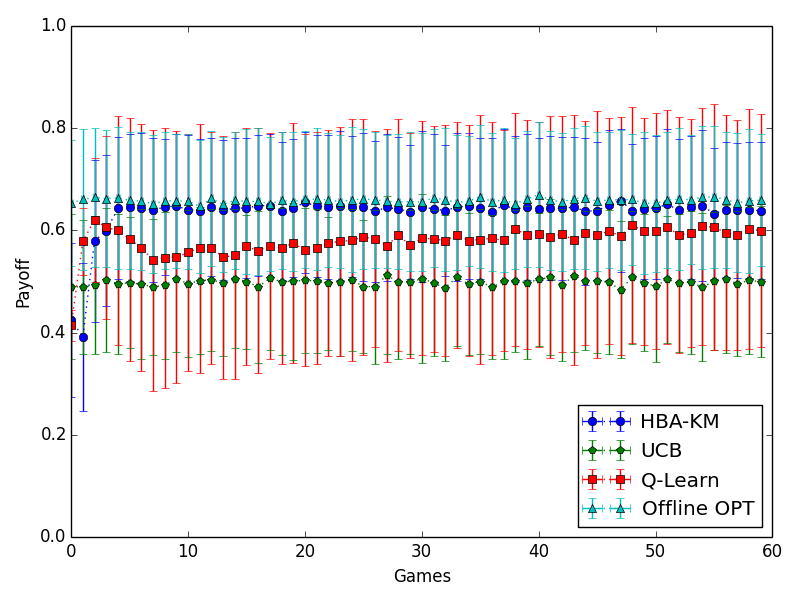}
        \caption{Exponential distribution}
\end{subfigure}
\caption{Revenue comparison and standard deviation against random strategies.}\label{fig:rand}
\end{figure}

In Figure \ref{fig:rand}, we can see the HBA-KM's performance against the randomised strategies of the advertiser. The performance of HBA-KM approximates the optimal offline benchmark, on every single occasion based on the distribution approximation that the subroutine \text{Random KM} performs.

As another metric we consider the average competitive ratio of each algorithm, when compared to the \emph{online optimal} one. The online optimal algorithm is the one with the best case cost; imagine that the publisher knows \emph{a priori} the sequence of bids the advertiser is going to play. Then he attains the online optimal policy by greedily selecting to sell the impression at the highest possible cost, up until the budget of the advertiser is exhausted. So for a bid $b$ and budget $B$, we have:
$$
\pi_{P}(b, B) = \max_{r \in \{0, 1\}} r \;\; s.t. \; r \leq b, \; r \leq B
$$

 The competitive ratio is defined as $\rho_{\textsc{Alg1}} = \frac{\textsc{Alg1}}{\textsc{Online-Opt}}$ and Table \ref{tab:cr} summarises the results over all the opposing strategies, adaptive and randomised respectively. The significant drop on the competitive ratio of every benchmark we witness when we move from facing adaptive strategies to randomised is expected, as the optimal online algorithm has knowledge of the exact sequence of bids, something powerful against truly random opponents.

\begin{table}[!ht]
\centering
\caption{Average Competitive Ratio, across all strategies and simulations}
\begin{tabular}{|l|c|c|c|c|}\hline
     & \multicolumn{2}{|c|}{\textit{Against Adaptive}} &\multicolumn{2}{|c|}{\textit{Against Randomised}}\\ \hline
\textbf{Algorithm Name}&\textbf{Competitive Ratio}&\textbf{Std}&\textbf{Competitive Ratio}&\textbf{Std}\\ \hline
\textsc{Offline-Opt} & 0.9721 & 0.0064 & 0.7657 & 0.1009\\ \hline
\textsc{HBA-KM} & 0.9245 & 0.2073& 0.7434 & 0.1585\\ \hline
\textsc{Q-Learn} & 0.7976 & 0.1650& 0.6165 & 0.2673\\ \hline
\textsc{UCB} & 0.7004 & 0.2334& 0.6218 & 0.1751\\
\hline\end{tabular}\label{tab:cr}
\end{table}
\vspace{-0.5cm}

\subsection{Neural Network agent}

So far our experiments only included opposing agents that are in the hypothesised type space $\Theta_A$ of our own algorithm. Unfortunately, this is not always the case in real life scenarios, as an agent in this marketplace should be able to face opposing strategies he cannot expect, or model explicitly, in real time. This is the question we sought to answer in the second part of our experiments; what happens when such an agent enters the market and is our algorithm still competitive against him?

We choose a Neural Network agent as our unknown opponent for two reasons. The first reason is that a NN does not belong to the hypothesised type space $\Theta_A$ of our own agent, therefore our type space should be descriptive enough to be able to model adequately such unknown agents on the fly. The second reason, consistent to the second part of our experiments, where we use a Neural Network trained in a mixture of the opposing publishers, is that we want to capture the inherent heterogeneity an agent faces in this market, where his opponents are trained against a variety of pricing algorithms. Here we simulate the Neural Network with up to 4 Hidden Layers and train it at each arriving impression. 

Our exact parameters for the simulation follow: Two input layers, the bid of the advertiser and the reserve price of the publisher. 1 up to 4 Hidden Layers. One output layer, the bidder's immediate payoff. Each node is fully-connected with nodes of next layer and we use a standard sigmoidal threshold function. We train online for every instance of the first day of simulations and for 100 impressions for each subsequent day. The network is trained until convergence at the end of each simulation day. The optimisation step is using a Hill Climber approach. 

\subsubsection{NN trained on a single opposing agent}

In the first set of experiments, we train the neural network using samples from his current opponent. The two input layers of the network are the bid and the reserve price for each impression that arrives and the single output is the revenue of the advertiser. We run the simulations using a neural network with 1 up to 4 hidden layers. Table \ref{tab:crnnet} summarises the competitive ratio of each algorithm against this agent.
\vspace{-0.5cm}
\begin{table}[!ht]
\centering
\caption{Average Competitive Ratio against a NN agent, across all simulations}
\begin{tabular}{|c|c|c|} \hline
\textbf{Algorithm Name}&\textbf{Competitive Ratio}&\textbf{Std}\\ \hline
\textsc{HBA-KM} & 0.9423 & 0.2697\\ \hline
\textsc{Q-Learn} & 0.8699 & 0.3470\\ \hline
\textsc{UCB} & 0.8469 & 0.4119\\
\hline\end{tabular}\label{tab:crnnet}
\end{table}
\vspace{-0.5cm}

\subsubsection{NN trained on a mixture of opposing agents}

In the second set of experiments, we train the neural network in a mixture of the opposing publishing agents. Specifically, for subsequent chunks of 100 impressions, the NN agent is trained with samples from the \textsc{HBA-KM}, the \textsc{UCB} and the \textsc{Q-Learn} respectively, throughout the first day of the simulations (1000 impressions). The reasoning behind this type of training is that in real world auctions we should expect to face algorithms trained in a variety of scenarios and, as such, we will not be able to model these agents explicitly. Table \ref{tab:crnnetmix} shows that against this opposing network, our algorithm stays highly competitive, even compared to the online optimal benchmark through the competitive ratio.
\vspace{-0.5cm}
\begin{table}[!ht]
\centering
\caption{Average Competitive Ratio against a mixed Neural Network agent, across all simulations}
\begin{tabular}{|c|c|c|} \hline
\textbf{Algorithm Name}&\textbf{Competitive Ratio}&\textbf{Std}\\ \hline
\textsc{HBA-KM} & 0.9501 & 0.2222\\ \hline
\textsc{Q-Learn} & 0.7957 & 0.4336\\ \hline
\textsc{UCB} & 0.8630 & 0.3897\\
\hline\end{tabular}\label{tab:crnnetmix}
\end{table}
\vspace{-0.5cm}

\section{Discussion and Conclusions}

In this paper, we address the learning problem faced by the publisher in an ad exchange, an interaction that is both practically significant and scientifically challenging. We propose the use of a novel methodology for learning in multiagent interactions, showing how this enables us to achieve substantial empirical improvements in simulations involving the TAC AdX domain.

Although we have not performed the theoretical analysis of these extensions, we conjecture that HBA-KM best response actions will always converge to an approximately optimal policy against either stochastic or deterministic opponents, within this posted price auction mechanism. This is based on the observation that the challenge is twofold, with each individual piece having known properties. The Random KM estimator can approximate successfully any given distribution, or a specific family of random distributions, while HBA converges to the correct beliefs over his hypothesised type space $\Theta_A$. 

A useful future direction would be to expand this to the case where there are multiple advertisers and a publisher interacting with each other in the ad exchange market. The main question here becomes whether there is a way (a) to model explicitly every single one of your opponents, or (b) to model the market price, i.e. the price $\rho^t$ that an agent $A \in \mathbf{A}$ faces in each step $t$ of the auction and is derived from the joint actions of every other agent in the auction $\mathbf{A}-\{A\}$. An algorithm that answers successfully either of those questions, would come a long way to us understanding the implicit interactions between different agent types, and will probably have other implications in situations where the modelling of your opponent, or teammate, on the fly is the core of the problem, such as the \emph{ad hoc teams} challenge \cite{stone2010ad}.

%
%

%
%
%
\bibliographystyle{splncs04}
\bibliography{mybibliography}
%




\end{document}